%% file: main.tex
\documentclass[12pt, a4paper]{extarticle}
\usepackage{geometry} 
\usepackage{courier}
\usepackage[bottom]{footmisc}
\usepackage{apacite}
\usepackage{natbib}
\usepackage{indentfirst}
\usepackage{amsmath}
\usepackage{amssymb}
\usepackage{amsthm}
\usepackage{graphicx}
\usepackage[colorlinks=true,linkcolor=blue,citecolor=blue,urlcolor=black]{hyperref}
\usepackage{titlesec}
\usepackage{pifont}
\usepackage{comment}
\usepackage{setspace}
\usepackage{bm}
\usepackage{enumerate}
\titlelabel{\thetitle.\,\,}
\newcommand{\citew}[1]{\citeauthor{#1}'s (\citeyear{#1})}
\newtheorem{theorem}{Theorem}

\newtheorem{lemma}{Lemma}
\newtheorem{assumption}{Assumption}

\newtheorem{assumptionalt}{Assumption}[assumption]

\newenvironment{assumptionp}[1]{
  \renewcommand\theassumptionalt{#1}
  \assumptionalt
}{\endassumptionalt}

\parskip=\baselineskip
\title{Double Robust Mass-Imputation with Matching Estimators}
\author{Ali Furkan Kalay \thanks{School of Economics, The University of Queensland, Australia. Email: \href{a.kalay@uq.edu.au}{a.kalay@uq.edu.au} }}

\date{\today}

\begin{document}

\maketitle

\begin{abstract}
    This paper proposes using a method named Double Score Matching (DSM) to do mass-imputation and presents an application to make inferences with a nonprobability sample. DSM is a $k$-Nearest Neighbors algorithm that uses two balance scores instead of covariates to reduce the dimension of the distance metric and thus to achieve a faster convergence rate. DSM mass-imputation and population inference are consistent if one of two balance score models is correctly specified. Simulation results show that the DSM performs better than recently developed double robust estimators when the data generating process has nonlinear confounders. The nonlinearity of the DGP is a major concern because it cannot be tested, and it leads to a violation of the assumptions required to achieve consistency. Even if the consistency of the DSM relies on the two modeling assumptions, it prevents bias from inflating under such cases because DSM is a semiparametric estimator. The confidence intervals are constructed using a wild bootstrapping approach. The proposed bootstrapping method generates valid confidence intervals as long as DSM is consistent. 
\end{abstract}



\newpage 

\section{Introduction}

This paper proposes a novel mass-imputation method to integrate two datasets which we label Double Score Matching (DSM). DSM is a consistent estimator as long as one of the two modeling assumptions is correct - this property is known as \textit{double robustness}. Even if the consistency assumptions fail, DSM could prevent bias from inflation under certain scenarios as matching estimators are semiparametric. The paper presents an application of DSM mass-imputation to make inferences with a sample not representative of the population (a nonprobability sample). 

Nonprobability samples are increasingly becoming the dominant type of data and becoming available for research purposes such as internet surveys, administrative records, and most types of big data. Various methods are available to make inferences with a nonprobability sample\footnote{e.g., \cite{elliott2017inference, buelens2018comparing, kim2020data}}. We mass-impute the variable of interest from a nonprobability sample to a population-representative sample (probability sample) using DSM. Our aim is to estimate the population mean using the mass-imputed data. DSM is a non-smooth estimator, and hence deriving its asymptotics is not straightforward. We provide consistency results and construct the confidence interval using a wild bootstrapping approach \citep{otsu2017bootstrap}. 

The Double Score Matching framework was proposed by \citet{long2012doubly} to impute missing values, and by \cite{antonelli2018doubly} to estimate average treatment effects with high-dimensional data. DSM is a nearest neighbor algorithm that uses the propensity\footnote{The propensity score is a probability measure for observations being in a certain sample, e.g., \textit{conditional probability of being in treatment sample}. \citet{rosenbaum1983central} showed that it can be used to balance the distributions of two samples (called \textit{balance score}).} and prognostic\footnote{The prognostic score is the condition outcome variable such as $E[y|\bm x]$. \citet{hansen2008prognostic} show that it could be used as an analogue of the propensity score to balance two sample distributions.} scores as distance metrics instead of a set of $k$ dimensional covariates. The use of two balance scores increases the convergence rate of the nearest neighbor algorithm by reducing the matching dimension from $k$ to 2. 

The proposed method directly relates to \citet{yang2018integration} where they use the nearest neighbor algorithm for mass-imputation. They find the nearest neighbors by calculating the distances (e.g., Euclidean Distance) between $k$ dimensional observations. However, the nearest neighbor algorithm does not perform well when $k$ is large, i.e., the convergence rate slumps sharply as the matching covariates' dimension increases. Specifically, the bias disappears at rate $O_p(N^{-1/k})$. Thus, mass-imputation with the nearest neighbor algorithm is only viable when the nonprobability sample is substantially large. The use of balance scores as a distance metric instead of covariates overcomes this problem. For example, the bias term with the propensity score matching (PSM) \citep{rosenbaum1983central} converges at rate $O_p(N^{-1})$, and hence it is negligible when constructing the confidence intervals. The propensity scores, alternatively, could be used as a weight to adjust distributions. The method is called Inverse Probability Weighting (IPW). However, extreme values of the propensity scores could lead the estimator to have a large variance\footnote{Example: If the estimated propensity score for observation $i,\, (\hat \pi _i )$, is small enough, $\frac{y_i}{\hat \pi _i}$ will inflate.}. Hence, IPW is very sensitive to the misspecification of the propensity score model. 

The propensity score matching could be useful to balance the control and treatment samples to estimate the Average Treatment Effect. However, its predictive power could be weak. Hence, Predictive Mean Matching (PMM) is more popular when predictive accuracy is required, e.g., missing value imputation \citep{little1988missing}. PMM has similar asymptotic properties to propensity score matching. An alternative is to use predictions of the outcome models as imputations - instead of using scores as a distance metric for matching. For example, \citet{kim2020combining} propose a model-based mass-imputation method to make inferences with a nonprobability sample; \citet{chen2020nonparametric} use kernel-smoothers and generalized additive models to do mass-imputation. 

\citet{robins1994estimation} proposed an estimator that combines mass-imputation and weighting called Augmented Inverse Probability Weighting. The AIPW is \textit{double robust} in the sense that as long as one of the two models is correctly specified, then the estimator is consistent. We refer to AIPW as Double Robust Estimator (DRE) following the more recent literature. DRE has widespread use along with various extensions for different purposes, such as making inferences with a nonprobability sample \citep{chen2020doubly} and missing value imputation \citep{carpenter2006comparison}. 

DSM is a nearest neighbor algorithm that calculates the matching distance with two balance scores (propensity and prognostic scores). DSM has similar asymptotic properties to PSM and PMM while it achieves better robustness properties. We consider DRE as a natural benchmark for our method because the two methods have similar consistency assumptions. We show that DSM is robust to functional form misspecifications, unlike DRE. 

The rest of the paper is organized as follows. Section \ref{MassImputation} presents the estimator and all consistency results (all proofs are provided in the Appendix). Section \ref{Application} shows how the estimator can be used to make inferences with a nonprobability sample.  Section \ref{Simulation} presents the simulations to demonstrate two features of the Double Score Matching method: double robustness and efficiency with nonlinear confounders.  Section \ref{Conclusions} provides brief conclusions. 

\section{Mass Imputation with Double Score Matching} \label{MassImputation}

We first establish notation. Suppose $ \mathcal{U} $ is an N-sized population with fixed population values $F_N = \{ (\bm x_i, y_i), i \in \mathcal{U}\}$. The population mean is $ \mu = N^{-1}\sum _{i \in \mathcal{U}}E[y_i|\bm x_i] $. Let the nonprobability ($S_A$) and probability samples ($S_B$) be drawn from the population $\mathcal{U}$ with the inclusion probabilities $\pi _i ^A$ and $\pi _i ^B$, respectively. Note that $\pi _i ^A$ (we refer to it as propensity score from now on) is not observable, but $\pi _i ^B$ is often available for the probability sample from sample weights, $d_i = 1/\pi _i ^B\,,i\in S_B$. Only the nonprobability sample contains the variable of interest, $y_i\in \mathbb{R}$. We assume that both samples contain the covariate vector $\bm x_i\in \mathbb{R}^k$. Consequently, we have a nonprobability sample such that $\{ (\bm x_i,y_i), i\in S_A \}$ with size $N_A$ and a probability sample such that $\{ (d_i,\bm x_i), i\in S_B \}$ with size $N_B$. 

Let $R_i$ be an indicator variable such that $R_i=1$ if observation $i$ is in the nonprobability sample, and $R_i=0$ otherwise. We define the propensity score as follows: $\pi _i ^A = Pr(R_i=1|\bm x_i,y_i)$. By definition of the inclusion probability $\pi _i ^B$ it must be that $\sum _{i\in S_B} d_i= \hat{N} \xrightarrow{p} N$ as $N_B\to\infty$. We consider $1/ N_B\sum _{i\in S_B} E[y_i|\bm x_i]=\mu _B$ as a subpopulation parameter as $y_i$ is not observable in $S_B$. 

There are two objectives of the paper. The first one is estimating the subpopulation parameters $\mu _B$ with DSM mass-imputation. The second one is inferring the population mean $\mu $ using the imputations. Note that the nonprobability sample mean is not a consistent estimator of the population mean because of the selection bias in $R_i$, thus $E\frac{1}{N_A} \sum _{i\in S_A} y_i $ is not a consistent estimator of the population mean ($\mu $).

\begin{assumption}\label{a:1}
    Let $\bm x$ be a random vector of continuous covariates distributed in $\mathbb{R}^k$ with compact and convex support $\mathbb{X}$, with a density bounded and bounded away from zero.
\end{assumption}

\begin{assumption}\label{a:strong-ignorability}
    For almost every $\bm x \in \mathbb{X}$,\begin{enumerate}[ {(}i{)}]
        \item Unconfoundedness: $R$ is independent of $y$ conditional on $\bm x$,
        \item Positivity: $\eta < Pr(R=1|\bm x) < 1-\eta $ for some constant $\eta >0$.
    \end{enumerate}
\end{assumption}

Assumption \ref{a:strong-ignorability} is also known as Missingness at Random (MAR). In particular, the unconfoundedness is crucial for our concern as it cannot be tested. 

\begin{assumption}\label{a:relative-sample-size}
    (i) $R_i$, $\{(y_i, \bm  x_i)\}_{i=1} ^N$ are independent draws from $y,\bm x|R$ for $r=0,1$. (ii) Sample sizes of $S_A$ and $S_B$ go to infinity at same rate: $N_B^a/N_A \to A$ where $a= 1$ and $0<A<\infty$.
\end{assumption}

Assumption \ref{a:relative-sample-size} introduces a crucial parameter for the asymptotic results. $a=1$ implies that sample sizes  $N_A$ and $N_B$ converge to infinity at the same rate\footnote{$a=1$ is the conventional assumption in the literature implying that the proportion of the sample sizes is constant when sample sizes go to infinity.}. $N_B$ converges faster than $N_A$ if $a<1$. However, it is not a reasonable assumption because representative samples (e.g., a survey with sample weights) are often more expensive to collect than nonprobability samples (e.g., big data\footnote{Most types of big data are nonprobability samples. Several studies discuss data integration using big data in the literature \citep{yang2018integration, kim2019sampling, kim2020data}}). This argument suggests that $a>1$ could be a plausible assumption under certain conditions, i.e., when the nonprobability sample is too large. We will consider the implications of the following alternative assumption in asymptotic results:

\begin{assumptionp}{\ref*{a:relative-sample-size}$'$}\label{a:relative-sample-size-alt}
    Assumption \ref{a:relative-sample-size} (i) holds and (ii) $N_B^a/N_A \to A$ where $a> 1$ and $0<A<\infty$.
\end{assumptionp}

\begin{assumption}\label{a:continuity}
    For $R=0,1$, $E[y|\bm x, R]$ and $Var(y|\bm x, R)$ are Lipschitz in $\mathbb{X}$.
\end{assumption}

Assumption \ref{a:continuity} implies that conditional mean and variance are continuous mappings.  

\subsection{Double Score Matching} \label{DSM}

We match $M$ observations in sample $S_A$ with each observation in sample $S_B$ with replacement. The \textit{best} $M$ matchings are selected based on the distance between the observations' matching scores $\bm Z = [f(\bm x;\theta _r ), g(\bm x;\theta _y) ]$ where $\bm Z$ is a $(N_A+N_B)\times 2 $ matrix consisting of propensity and prognostic scores, respectively, 

\begin{align}
      f(\bm x;\theta _r ) &= P(R=1|\bm x),\\
      g(\bm x;\theta _y) &= E[y|\bm x].
\end{align}

$\theta _r$ and $\theta _y$ are estimated as follows:

\begin{align*}
    \hat \theta _r &= \max _{\theta _r} \quad \sum_{i \in \mathcal{S}_{{A}}} \log \left\{\dfrac{f(x_i;\theta _r )}{1-f(x_i;\theta _r )}\right\}+\sum_{i \in \mathcal{S}_{{B}}} d_{i}^{{B}} \log \left\{1-f(x_i;\theta _r )\right\},\\
    \hat \theta _y &= \min _{\theta _y } \quad \sum _{i \in S_A} \left (y_i - g(x_i;\theta _y)\right )  ^2.
\end{align*}

The conventional functional choices for propensity and prognostic scores are logistic and linear regression, respectively. The propensity score log-likelihood function has a Horvitz-Thompson estimator for the population parameter following \citet{chen2020doubly}. They showed that estimating the propensity scores with pooled data (a log-likelihood function without sample weights\footnote{Pooled data log-likelihood function:

$$ \sum_{i \in \mathcal{S}_{{A}}} \log \left\{\dfrac{f(x_i;\theta _r )}{1-f(x_i;\theta _r )}\right\}+\sum_{i \in \mathcal{S}_{{B}}} \log \left\{1-f(x_i;\theta _r )\right\}.$$}) leads to biased results since the true log-likelihood function must be estimated with the entire population. 

Let $\bm {\hat Z}$ be the score matrix that is constructed with estimated model parameters ($\hat \theta = [\hat \theta _r, \hat  \theta _y]$) and normalized by scores' standard deviation such that,

\begin{equation*}
    \bm {\hat Z} = \left [ \dfrac{f(\bm x; \hat \theta _r )}{\text{SD}(f(\bm x; \hat \theta _r ))}, \dfrac{g(\bm x;\hat \theta _y)}{\text{SD}(g(\bm x; \hat \theta _y))} \right ].
\end{equation*}

Let $\bm { Z}$ be the score matrix that is constructed with known model parameters ($\tilde \theta = [ \tilde \theta _r,   \tilde \theta _y]$) and normalized by scores' standard deviation such that,

\begin{equation*}
    \bm{ Z} = \left [ \dfrac{f(\bm x; \tilde \theta _r )}{\text{SD}(f(\bm x; \tilde \theta _r ))}, \dfrac{g(\bm x;\tilde \theta _y)}{\text{SD}(g(\bm x; \tilde \theta _y))} \right ],
\end{equation*}

Where parameters $\tilde \theta _r$ and $\tilde \theta _y$ are defined as follows: 

\begin{align*}
    \tilde \theta _r &= \max _{\theta _r} \quad E\left [ \sum_{i \in \mathcal{S}_{{A}}} \log \left\{\dfrac{f(x_i;\theta _r )}{1-f(x_i;\theta _r )}\right\}+\sum_{i \in \mathcal{U} } \log \left\{1-f(x_i;\theta _r )\right\}\right ]\quad \text{ and }\\
    \tilde \theta _y &= \min _{\theta _y } \quad E\left [ \sum _{i \in S_A} \left (y_i - g(x_i;\theta _y)\right )  ^2 \right ].
\end{align*}

Let $\bm z_i$ be the score vector $i$ in $\bm Z$, $\bm{\hat z_i}$ be the score vector $i$ in $\bm{\hat{Z}}$, and $j_m(i; \hat \theta )$ be the $m^{th}$ closest unit in $S_A$ to $i\in S_B$:

\begin{equation*}
    \sum_{l \in S_A} I\left\{\left\|\bm{\hat z_{l}}-\bm{\hat z_{i}}\right\| \leq\left\|\bm{\hat z} _{j_m(i; \hat \theta )}-\bm{\hat z}_{i}\right\|\right\}=m,
\end{equation*}

Where $I\{.\}$ is an indicator function and $||.||$ is the standard Euclidean norm. $j_m(i; \hat \theta )$ is the best $m^{th}$ match for observation $i$. Thus, the set of the best $M$ matches for $i\in S_B$ could be defined as follows:

\begin{equation*}
    \mathcal{J}_M(i;\hat \theta ) = \{j_1(i;\hat \theta ), ..., j_M(i;\hat \theta )\}.
\end{equation*}

The best matches are chosen using the estimated score matrix. $\bm x_i$ is assumed to be drawn from a continuous distribution, and therefore estimated scores are unique by assumption \ref{a:continuity}. Since we match with replacement, some observations from the nonprobability sample could be matched more than once, while others may not be matched at all. The distribution of the number of matches plays a vital role in variance estimates \citep{abadie2006large, abadie2008failure, otsu2017bootstrap}. Let $K_M(i; \theta)$ be the number of times that observation $i\in S_A$ is matched with the probability sample observations when scores are estimated with parameter $\theta $. 

\begin{equation}
    K_M(i; \hat \theta ) = \sum _{l \in S_B} I\{ i \in \mathcal{J}_M(l; \hat \theta )\} \qquad \text{ where } i\in S_A \label{eq:nmatchings}.
\end{equation}

DSM imputes the average of the best $M$ matches to the probability sample for all $i\in S_B$. After that, it is straightforward to estimate $\mu _B$: 

\begin{equation}
     \mu _{B} (\hat \theta ) = \dfrac{1}{N_B} \sum _{i \in S_B} \dfrac{1}{M} \sum _{j \in S_A} I\{ j \in \mathcal{J}_M(i ; \hat \theta ) \} y_j.\label{eq:dsm_imp}
\end{equation}

The mean is estimated by imputing the missing values $y_i$ in $S_B$ by:
\begin{equation}
    y_i (\hat \theta ) = \hat y_i = \dfrac{1}{M} \sum _{j \in \mathcal{J}_M(i; \hat \theta )} y_j.\label{eq:imputation}
\end{equation}

We also define imputation with the known scores for observation $i$ as $\tilde y_i = y_i (\tilde \theta )$. Equation \eqref{eq:dsm_asy} is equivalent to equation \eqref{eq:dsm_imp}; however, it will prove more convenient for the discussion on the estimation of the variance. 

\begin{equation}
    \mu _{B} (\hat \theta )  = \dfrac{1}{N_B} \sum _{i \in S_A} \dfrac{K_M(i; \hat \theta ) }{M} y_i. \label{eq:dsm_asy}
\end{equation}

DSM mass-imputation steps can be summarized as follows:

\begin{description}
    \item[Step-1] Estimate propensity score using samples $S_A$ and $S_B$; and prognostic score with sample $S_A$.
    \item[Step-2] Define the score matrix, $\bm Z$, using two normalized (by scores' standard deviations) scores. 
    \item[Step-3] Match each observation in the probability sample, $S_B$, with $M$ observations from the nonprobability sample, $S_A$, using the nearest neighbor algorithm.
    \item[Step-4] Impute all $i \in S_B$ with the average of the matchings. 
    \item[Step-5] Estimate the population mean by using imputed values.
\end{description}

\subsection{Asymptotic Results}

This section establishes the consistency results for DSM mass-imputation. DSM uses two balance score models to generate matching distance metrics. Thus the consistency of the estimator requires the following assumption:

\begin{assumption}\label{a:dr}
    Either propensity score model or prognostic score model is a valid balance score as in \citet{rosenbaum1983central} and \citet{hansen2008prognostic}, respectively. In other words:
    \begin{gather*}
        P(R=1|\bm x) = f(\bm x;\tilde \theta _r) \qquad \text{ or } \qquad E[y|\bm x] = g(\bm x;\tilde \theta _y).
    \end{gather*}
\end{assumption}

We say that model specification is a correct specification of the true score models if it is a valid balance score. This assumption could fail, for instance, if the data generating process has a nonlinear functional form unknown to the researcher. This scenario is simulated in section \ref{sec:nonlinear-sim} (and in appendix \ref{Appendix:simulation}).

\begin{lemma}
    \citep[Theorem 1]{antonelli2018doubly} Suppose assumptions 1-5 hold. Let $f (\bm x, \tilde \theta _r)$ be the true propensity score, $g (\bm x, \tilde \theta _y)$ be the true prognostic score, and let $h(\bm x, \tilde \theta)$ be an arbitrary function of $\bm x$.Then:

    \begin{equation*}
        y \perp R \,|\, f (\bm x, \tilde \theta _r), h(\bm x, \tilde \theta) \quad \text{ and } \quad y \perp R \,|\, g (\bm x, \tilde \theta _r), h(\bm x, \tilde \theta) .
    \end{equation*} \label{lemma:dr}
\end{lemma}

Lemma \ref{lemma:dr} implies the double robustness, i.e., any arbitrary function of $\bm x$ does not disturb the conditional independence provided that either the propensity or prognostic score is correctly specified \footnote{Note that lemma \ref{lemma:dr} is a special case of Theorem 1 in \citet{antonelli2018doubly} because we do not control for the treatment variable.}. Specifically,

\begin{align*}
    \mu _B &= \sum _{i\in S_B} E[y_i\,|\,f (\bm x _i, \tilde \theta _r), h(\bm x _i, \tilde \theta)]\qquad \text{ and }\\
    \mu _B &=  \sum _{i\in S_B} E[y_i\,|\,g (\bm x _i, \tilde \theta _y), h(\bm x _i, \tilde \theta)].
\end{align*}

\begin{lemma}
    Suppose assumptions 1-5 hold, then as $N_A\to \infty$, 

    \begin{equation*}
        E[\tilde y _i - E[y_i|\bm z_i]] = 0.
    \end{equation*}
\label{lemma:dr-specific}
\end{lemma}

\begin{proof}
    See appendix.
\end{proof}

Lemma \ref{lemma:dr-specific} is a weak result as scores are assumed to be known. It shows that each imputation with know scores is doubly robust, but it does not explain the asymptotic properties of the DSM. 

Following \citet{abadie2006large} we find the convergence rate of the DSM estimated with known scores ($\mu _B (\tilde \theta )$) by decomposing $\mu _B (\tilde \theta ) - \mu _B$ as follows:

\begin{equation}
    \mu _B (\tilde \theta ) - \mu _B = (\overline{\mu_B(Z)} - \mu _B) + E_M + B_M, \label{eq:atet}
\end{equation}

where 

\begin{align}
    \overline{\mu_B(Z)} - \mu _B&= \left \{\dfrac{1}{N_B} \sum _{i \in S_B}E[y_i|\bm z_i]\right \} - \mu _B,\label{eq:het}\\
    E_M &= \dfrac{1}{N_B} \sum _{i\in S_A} \dfrac{K_M(i; \tilde \theta )}{M}\epsilon _i,\label{eq:res}\\
    B_M &= \dfrac{1}{N_B} \sum _{i \in S_B} \dfrac{1}{M} \sum _{m=1}^M (E[y_i|\bm z_i] - E[y_i|\bm z_{j_m(i\ \tilde \theta )}]),\label{eq:bias}\\
    y_i &= E[y_i|\bm z_i] + \epsilon _i.
\end{align}

Equation \eqref{eq:atet} has the same asymptotic properties as the Average Treatment Effect of the Treated estimator (ATET) in \citet{abadie2006large}. Here we define the treatment variable such that $W_i = I\{ i\in S_B \}$. Assume that the nonprobability sample is the control sample and the probability sample is the treatment sample. The only difference between ATET and mass-imputation is that ATET is known to be zero. There is a selection bias in the nonprobability sample; hence ATET may not be estimated as zero. However, a non-zero ATET can only be estimated if assumption \ref{a:strong-ignorability} or \ref{a:dr} fails.

By Theorem 2 of \citet{abadie2006large}, $B_M$ converges at rate $o_p(N_B^{-a/2})$. It will be dominated by other terms if $a>1$ as $(\overline{\mu_B(Z)} - \mu _B) + E_M=O_p(N^{-1/2}_B)$. If $a\leq 1$, however, the $B_M$ is not dominated\footnote{Furthermore, $B_M$ dominates other terms if $a<1$.} and, hence, the variance of the bias term is not negligible in the estimation. 

\begin{lemma}\label{lemma:convergenceinprob}
Suppose assumptions 1-5 hold. Then, 

    \begin{equation*}
        \mu _B (\tilde \theta ) - \mu _B = O_p(N_B^{-1/2}).
    \end{equation*}
\end{lemma}

\begin{proof}
    See appendix.
\end{proof}

The proof of Lemma \ref{lemma:convergenceinprob} shows that Theorem 2 of \citet{abadie2006large} is applicable to equation \eqref{eq:atet}, i.e. equation \eqref{eq:atet} has the same asymptotic properties as the ATET estimator. The proof exploits the fact that ATET is known be zero as there is no ``treatment'' in sample $S_B$.

\citet{abadie2006large} do matching on known covariates. Hence, Lemma \ref{lemma:convergenceinprob} provides a weak result for DSM as it implicitly assumes propensity and prognostic scores are known. The following theorem replaces known scores $\bm Z$ with estimated scores $\bm{\hat Z}$ and shows that DSM with estimated scores converges at rate $O_p(N_B^{-1/2})$.

\begin{theorem}\label{theorem:strongproof}
    Suppose assumptions 1-5 hold. Then, 
    \begin{equation*}
        \mu_B (\hat \theta )   - \mu _B = O_p(N_B^{-1/2}).
    \end{equation*}
\end{theorem}

\begin{proof}
    See appendix.
\end{proof}

Theorem \ref{theorem:strongproof} shows that DSM mass-imputation converges at same rate when scores are known (Lemma \ref{lemma:convergenceinprob}) and estimated. The proof method follows \citet{antonelli2018doubly} with some modifications. The variance of the $\mu _B (\hat \theta)$ is discussed next. 

\subsection{Variance} 

\subsubsection{Variance of the Double Score Matching Estimator}

Let the variance of equation \eqref{eq:het} be $V^{\mu (Z)}$; the variance of equation \eqref{eq:res} be $V^{E}$; and the variance of equation \eqref{eq:bias} be $V^{B_M}$. The first is the variance of the scores' heterogeneity, the second is the variance conditional on scores, and the last is the variance of the bias term. Ignoring the bias term's variance, the variance of the matching estimator is the sum of the two terms, $V = E[V^{\mu (Z)}] + E[V^{E}]$. The bias term's variance is negligible under certain conditions, which we discuss later.  

\textit{Variance of Heterogeneity}: We use the imputations $\tilde y_i$ to estimate heterogeneity variance as follows:

\begin{equation}
    E[(\tilde{y} _i - \mu _B)^2] \simeq V ^{\mu (Z)}+ E\left [ \dfrac{1}{M^2} \sum _{j\in J_M(i; \tilde \theta )} \epsilon _j ^2 \right ]. \label{eq:var_extra}
\end{equation}The LHS could be estimated as $\sum _{i \in S_B} (\hat y_i -  \mu _B(\hat \theta)) ^2$ and the second term on the RHS could be estimated as follows:

\begin{equation}
    \dfrac{1}{N_B^2} \sum _{j\in S_B} E \left [ \dfrac{1}{M^2} \sum _{j\in J_M(i; \tilde \theta )} \epsilon _j ^2 \Big | \bm Z \right] \simeq \dfrac{1}{N_B^2} \sum _{i\in S_A} \left ( \dfrac{K_M(i;\tilde \theta )}{M^2}  \right) \sigma ^2 (\bm z_i ) \label{eq:var_adj},
\end{equation}

where $\sigma (\bm z_i) = Var (y_i|\bm z_i)$. We can estimate the $V^{\mu (Z)}$ by taking the difference of equations \eqref{eq:var_extra} and \eqref{eq:var_adj}. Note that $\sigma (\bm z_i)$ is unknown yet. 

\begin{equation}
    \tilde V ^{\mu (Z)} = \dfrac{1}{N_B^2} \sum _{i\in S_B} (\tilde y_i - \mu _B (\tilde \theta)) ^2  - \dfrac{1}{N_B^2} \sum _{i\in S_A} \left ( \dfrac{K_M(i;\tilde \theta )}{M^2}  \right) \sigma ^2 (\bm z_i ).
\end{equation}

\textit{Conditional Variance}: Assuming that $K_M(i; \tilde \theta )$ is deterministic, the conditional variance is as follows:

\begin{equation}
    V^E = \dfrac{1}{N_B^2} \sum _{i\in S_A} \left ( \dfrac{K_M(i; \tilde \theta )}{M} \right ) ^2 \sigma ^2 (\bm z_i).
\end{equation}

Both $\tilde V ^{\mu (Z)}$ and $V^E$ require the estimation of $\sigma (\bm z_i) $. \citet{abadie2006large} proposes the following method to estimate it:

\begin{equation}
    \tilde \sigma ^2 (\bm z_i) = \dfrac{J}{J+1} \left ( y_i - \dfrac{1}{J} \sum _{m=1} ^J y_{\mathit{l}_m (i;\tilde \theta)} \right )^2, \label{eq:residualvar1}
\end{equation}

where $J$ is a fixed term, ${\mathit{l}_m (i;\tilde \theta)}$ is the $m^{th}$ closest unit to $i\in S_A$ in the sample $S_A$. In other words, we match all $i\in S_A$ to the closest $J$ observations in $S_A$. The variance estimator is only asymptotically unbiased\footnote{If error terms are known or expected to be homoskedastic, $\sigma ^2 (\bm z_i)$ could be replaced with the following to make computation easier:

\begin{equation}
    \tilde \sigma ^2= \dfrac{1}{N_B} \sum _{i \in S_B} \tilde \sigma ^2 (\bm z_i), \label{eq:hom_error}
\end{equation}

In addition, using equation \eqref{eq:hom_error} could be more efficient in such cases.} and $J$ is often chosen larger than $M$. 

Combining previous equations, we estimate the variance of $\mu _B (\hat \theta) $ as follows:

\begin{equation}
    \tilde V = \dfrac{1}{N_B} \sum _{i\in S_B} (\tilde y_i - \mu _B (\tilde \theta)) ^2 + \dfrac{1}{N_B} \sum _{i \in S_A } \dfrac{K_M(i;\tilde \theta ) (K_M(i;\tilde \theta) -1)}{M^2} \tilde \sigma ^2 (\bm z_i). \label{eq:var}
\end{equation}

\citet{abadie2016matching} showed that the variance of matching estimator with estimated propensity scores differs from variance with known propensity scores. They derived the adjustment terms for ATE and ATET. \citet{yang2020asymptotic} presented a similar result for predictive mean matching that asymptotic results of matching on the known and estimated prognostic score are different. \citet{yang2020double} establish the martingale central limit theorem for DSM with estimated scores. The adjustment terms in asymptotic results with estimated scores are very complex to estimate. \citet{yang2020double}, for instance, construct the DSM's ATE confidence intervals with bootstrapping; \citet{yang2020asymptotic} estimate the variance of predictive mean matching with bootstrapping. Both use a similar methodology proposed by \citet{otsu2017bootstrap} which is discussed in the next section. 

\subsubsection{Wild Bootstrapping}

\citet{abadie2008failure} demonstrated that the naive bootstrapping approach\footnote{Using the matching estimator independently for each bootstrap sample.} fails with matching estimators. The main reason is that bootstrap sampling cannot preserve the distribution of $K_M(i;\tilde{ \theta })$. \citet{otsu2017bootstrap} proposed another method that re-samples the residuals. They, technically, generate bootstrap samples without re-estimating the $K_M(i;\tilde \theta)$. In other words, the number of times observation $i$ used for matching is considered to be a characteristic of the observation. 

\citet{yang2020double} and \citet{yang2020asymptotic} also adopted a similar approach to \citet{otsu2017bootstrap} due to complexity of estimating the variance adjustment term for the estimated scores. They proposed a parallel bootstrapping method that captures the uncertainty of the scores via a de-biasing term. 

We adopt the wild bootstrapping method proposed by \citet{otsu2017bootstrap} with \citew{mammen1993bootstrap} two point weight distribution. 

\begin{description}
    \item[Step-1] Draw bootstrap sample weights, $\{ w_i ^{(b)}\} _{b=1} ^B$, where $w_i ^{(b)}$ is independently drawn for each $i$ and $b$ from the following probability distribution:
    \begin{equation*}
        w_i ^{(b)} = \begin{cases}
            -(\sqrt{5} - 1)/2& \text{ with probability  } \quad (\sqrt{5} + 1)/2\sqrt{5}\\
            (\sqrt{5} + 1)/2 &\text{ with probability  } \quad (\sqrt{5} - 1)/2\sqrt{5}
        \end{cases}
    \end{equation*}  
    \item[Step-2] Compute the bootstrap $b$ residual:
    \begin{equation*}
        \hat q^{(b)} = \dfrac{1}{N_B} \sum _{i\in S_A} w_i ^{(b)} \left ( \dfrac{K_M(i;\hat \theta ) \, \left (y_i - \mu _B (\hat \theta )\right )}{M}  \right )
    \end{equation*} 
    \item[Step-3] Repeat steps 1-2 $B$ times. 
    \item[Step-4] Let $\hat q_{a}$ be the $a^{th}$ quantile of $\hat  q^{(b)}$. The $100(1-a) \% $ bootstrap confidence interval is:
    \begin{equation*}
        [\mu _B (\hat \theta) - \hat q_{1-a/2} ,\, \mu _B (\hat \theta) - \hat q_{a/2} ]
    \end{equation*} 
\end{description}

Both equation \eqref{eq:var} and our proposed bootstrapping method neglect the variance of the bias terms. However, the bias term is not negligible with assumptions 1-5 because the bias term \eqref{eq:bias} converges at same rate with equations \eqref{eq:het} and \eqref{eq:res}. However, it is negligible if the sample $S_A$ goes to infinity at a higher order than sample $S_B$. In other words, we can construct valid confidence intervals if $a>1$ if assumption \ref{a:relative-sample-size-alt} holds. Assumption \ref{a:relative-sample-size-alt} is not necessarily stronger than \ref{a:relative-sample-size}; and could be weaker if sample $S_A$ is large enough and if it is expected to go to infinity at a higher rate than sample $S_B$. 

Alternatively, we can predict the bias term based on our modeling assumption (prognostic score). The following section discusses how to de-bias the DSM to construct valid confidence intervals.

\subsubsection{De-Biased Estimator}

\citet{abadie2011bias} showed that bias term would converge faster than $O_p(N_B ^{-a/2})$ if the matching estimator is de-biased with a modeling assumption. It converges faster than usual m-estimators because as sample size increases, matching discrepancy reduces, and m-estimator converges simultaneously. Recall equation \eqref{eq:bias} where the bias is the difference between conditional outcomes of observation $i$ and its matching units. We already make modeling assumptions on conditional outcomes (prognostic score); hence we can  predict the bias as follows:

\begin{equation}
    \hat B_M = \dfrac{1}{N_B} \sum _{i \in S_B} \dfrac{1}{M} \sum _{m=1}^M (g(\bm x_{j_m(i;\hat \theta )}; \hat \theta _y) - g(\bm x_i;\hat \theta _y)).
\end{equation}

The resulting de-biased estimator will be:

\begin{equation}
    \mu _B^d (\hat \theta ) = \mu _B (\hat \theta ) - \hat B_M.\label{eq:de-biased}
\end{equation}

De-biasing the DSM makes the bias term's variance negligible; hence valid confidence intervals can be constructed. Even if the prognostics score model is misspecified, DSM is still asymptotically consistent (conditional on propensity score is correctly specified) because the matching discrepancies disappear at rate $O_p(N_B^{-1/2}) $. Following \citet{otsu2017bootstrap}, we will consider the $K_M (i;\hat \theta )$ as a characteristic of observation $i\in S_A$. Equation \ref{eq:de-biased} is rewritten as follows:

\begin{equation}
    \mu _B^d (\hat \theta ) = \dfrac{1}{N_B} \sum _{i\in S_A} \left ( \dfrac{{K}_M(i;\hat \theta ) (y_i - g(\bm x_i; \hat \theta _y))}{M} \right ) + \dfrac{1}{N_B} \sum _{i\in S_B} g(\bm x_i;\hat \theta _y) .
\end{equation}

The proposed wild bootstrapping could be implemented with the de-biased estimator by using equation \eqref{eq:de-biased} and changing the step-2 as follows:

\begin{description}
    \item[Step 2$'$:] Compute the bootstrap $b$ residual:
    \begin{equation*}
        \hat q^{(b)} = \dfrac{1}{N_B} \sum _{i\in S_A} w_i ^{(b)} \left ( \dfrac{{K}_M(i;\hat \theta ) (y_i - g(\bm x_i; \hat \theta _y))}{M} \right ) + \dfrac{1}{N_B} \sum _{i\in S_B} w_i ^{(b)} \left ( g(\bm x_i;\hat \theta _y) - \mu _B ^{d} (\hat \theta ) \right ).
    \end{equation*}
\end{description}

\section{Making Inferences with a Nonprobability Sample} \label{Application}

The previous section showed that the probability sample mean is estimated consistently. Now, we use DSM mass-imputation to make inferences with a nonprobability sample. We employ the Hajek Estimator \citep{hajek1964asymptotic} to find the sample mean using the imputations:

\begin{equation}
    \hat y_i = \dfrac{1}{M} \sum _{j \in \mathcal{J}_M(i; \hat \theta )} y_j .\label{eq:prediction}
\end{equation}

The population mean is estimated with the Hajek estimator as follows:

\begin{equation}
    \mu _{\Psi} = \dfrac{1}{\sum _{i\in S_B} d_i} \sum _{i\in S_B} d_i y_i = \dfrac{1}{\hat N} \sum _{i\in S_B} d_i y_i.
\end{equation}

We replace the $y_i$ with equation \eqref{eq:prediction} as $y_i$ are not observable:

\begin{align*}
     \mu _{DSM} (\hat \theta ) &=  \dfrac{1}{\hat N}  \sum _{i\in S_B} d_i \dfrac{1}{M} \sum _{j \in \mathcal{J}_M(i; \hat \theta )} y_j\\
    &=  \dfrac{1}{\hat N}  \sum _{i\in S_B} d_i \hat y_i\\
    &= \dfrac{1}{\hat N}  \sum _{i\in S_B} d_i  y_i - \dfrac{1}{\hat N}  \sum _{i\in S_B} d_i \hat \epsilon_i\\
    &=  \mu _{\Psi } - \dfrac{1}{\hat N}  \sum _{i\in S_B} d_i \hat \epsilon_i,
\end{align*}

where $\hat \epsilon_i$ is the residual of the DSM estimator, $y_i = \hat y_i + \hat \epsilon_i$.

Note that if the weighted mean of the error converges to zero, it implies $ \mu _{DSM} (\hat \theta )$ converges to the true population mean, $\mu$. 

\begin{theorem}
Suppose assumptions 1-5 hold, and $d_i \perp \epsilon _i| \bm z_i$. Then,

\begin{equation*}
    \mu _{DSM} (\hat \theta ) - \mu = O_p(N_B ^{-1/2}).
\end{equation*}

\end{theorem}

\begin{proof}
    See appendix.
\end{proof}

The estimation of the variance is not straightforward as we have a sample with unequal weights. The variance estimates require the knowledge of joint inclusion probabilities ($\pi ^B _{ij}$) as well as inclusion probabilities ($\pi _i ^B$) \citep{hajek1964asymptotic}. Although the latter is usually available with surveys, the joint inclusion probabilities are not.

Various methods are available to estimate the variance of a sample with unequal weights by approximating the joint inclusion probabilities or by bootstrapping/jackknife estimators\footnote{e.g., \citet{escobar2013new, berger2007jackknife}}. Nevertheless, the existing methods could not be facilitated directly as $y_i$ is unknown. Hence, we extend the proposed wild bootstrapping method with unequal sample weights for the de-biased DSM. 

\begin{align}
    \mu _{DSM} ^d (\hat \theta ) = \mu _{DSM}  (\hat \theta ) - \hat B_M ^d,
\end{align}

where 

\begin{equation}
    \hat B_M ^d = \dfrac{1}{\hat N} \sum _{i \in S_B}d_i \dfrac{1}{M} \sum _{m=1}^M (g(\bm x_i;\hat \theta _y) - g(\bm x_{j_m(i; \hat \theta )}; \hat \theta _y)).
\end{equation}

The sample weights require adjustments in bootstrapping Step-2 as follows: 

\begin{description}
    \item[Step 2$''$:] Compute the bootstrapped residual:
    \begin{equation*}
        \hat q^{(b)} = \dfrac{1}{\hat N} \sum _{i\in S_A} w_i ^{(b)} \left ( \dfrac{\bar{K}_M(i;\hat \theta ) (y_i - g(\bm x_i; \hat \theta _y))}{M} \right ) + \dfrac{1}{\hat N} \sum _{i\in S_B} w_i ^{(b)} d_i \left ( g(\bm x_i;\hat \theta _y) - \mu _{DSM} ^{d} (\hat \theta ) \right )
    \end{equation*}

    $\bar{K}_M(i;\hat \theta )$ is sum of the sample weights of observations where $i\in S_A$ used for matching. For example, if observation $i\in S_A$ used two times as matching for observations $j_1,\, j_2 \in S_B$, then $\bar{K}_M(i;\hat \theta ) = d_{j_1}+ d_{j_2}$.
\end{description}

\section{Simulation} \label{Simulation}

We simulate two samples $S_A$ and $S_B$ where sample $S_A$ is a nonprobability sample and sample $S_B$ is a probability sample. Both samples are drawn from a superpopulation ($F_N$) with Poisson sampling and probability proportional to size (PPS) sampling, respectively. We estimate two parameters: (1) The probability sample mean $\mu _B = \sum _{i\in S_B} E[y_i|\bm x_i]$ and (2) Population mean $\mu = \sum _{i\in F_N} E[y_i|\bm x_i]$.

We conduct two simulations. The first one replicates the DGP used by \citet{chen2020doubly} where the data generating process (DGP) has linear confounders. This is, then, extended to include nonlinear confounders.  The last section presents a number of simulations for the proposed wild bootstrapping method.

\subsection{Simulation with Linear Confounders}\label{sec:linearvar}

This section presents a simulation that replicates that in \citet{chen2020doubly}. Suppose that the variable of interest, $y$, is generated by the following stochastic process:

\begin{equation}
    y_{i} = 2 + x_{1i} + x_{2i} + x_{3i} + x_{4i} + \sigma \epsilon _i .\label{eq:outcomemodel}
\end{equation}

The DGP has four confounders which are drawn from the following distributions:

\begin{align*}
    x_{1i} &= z_{1i}\\
    x_{2i} &= z_{2i} + 0.3x_{1i}\\
    x_{3i} &= z_{3i} + 0.2(x_{1i}+ x_{2i})\\
    x_{4i} &= z_{4i} + 0.1(x_{1i}+ x_{2i} + x_{3i}) \\
    z_{1i} &\sim bernoulli(0.5)\\
    z_{2i} &\sim uniform(0,2)\\
    z_{3i} &\sim exp(1) \\
    z_{4i} &\sim\chi ^2(4)\\
    \epsilon _i&\sim N(0,1) \quad \text{(i.i.d.)}
\end{align*}

The value of $\sigma $ is chosen such that the correlation between $y$ and $X'\beta$ is 0.3, which is shown by the parameter $\rho=0.3$. 

The DGP for the propensity score, $\pi _i ^A$, is as follows:

\begin{equation}
    \log \left ( \dfrac{\pi ^A _i}{ 1-\pi ^A _i}\right ) = \theta _0 + 0.1 x_{1i}+ 0.2 x_{2i}+ 0.1 x_{3i}+ 0.2 x_{4i} \label{eq:propensityscoremodel}
\end{equation}

$\theta _0$ is selected such that $\sum _{i=1} ^N \pi ^A _i=N_A=500$ where $N=20000$ is the population size and $N_A$ is the nonprobability sample size\footnote{We allow for variable sample sizes for the nonprobability samples. $N_A$ is the expected sample size for Poisson sampling. }. The nonprobability sample ($S_A$) is drawn by Poisson sampling using the propensity scores as inclusion probabilities. The probability sample with size $N_B=1000$ is drawn by PPS sampling using inclusion probabilities $\pi ^B _i$ that is proportional to $p_i = c + x_{3i}$. $c$ is chosen such that $\max \{ p_i \} / \min \{ p_i \} = 50$ and and propensity scores are adjusted such that $\sum _{i=1} ^N \pi ^B _i=N_B$. 

We evaluate four different scenarios for the simulation: (TT) Both the propensity, and prognostic score models are correctly specified; (FT) The prognostic score model is misspecified, but the propensity score model is correctly specified; (TF) The prognostic score model is correctly specified, but the propensity score model is misspecified; (FF) Both models are misspecified. The misspecified models omit the variable $x_{3i}$ in either equation \eqref{eq:outcomemodel} or \eqref{eq:propensityscoremodel}.

$ \mu _B ^{(s)} (\hat \theta)$ is the sample $S_B^{(s)}$ mean estimate in simulation $s$. We compare it with the population parameter in the same simulation, $\mu _B ^{(s)} = \frac{1}{N_B}\sum _{i=S_B^{(s)}}  E[y_i|\bm x_i]$, in terms of relative bias and mean squared error. 

\begin{align*}
    RB &= \dfrac{1}{S}\sum _{s=1} ^S\dfrac{\mu _B ^{(s)} (\hat \theta) - \mu _B ^{(s)}}{\mu _B ^{(s)}} \times 100,\\
    MSE &= \dfrac{1}{S}\sum _{b=1} ^S\left (\mu _B ^{(s)} (\hat \theta) - \mu _B ^{(s)} \right ) ^2.
\end{align*}

\input{tables/sample S_B table/sampleb.tex}

We set $S=2000$ for all simulations and $M=3$ for both DSM and de-biased DSM. Table \ref{tab:sampleB} shows sample $S_B$ mean estimation results. Both DSM and De-biased DSM converge to the true sample $S_B$ mean as long as one of the score models is correctly specified. The results also reveal a possible disadvantage of using the de-biased estimator. If the prognostic score is misspecified, relative bias (and MSE) could be greater. The risk of using the de-biased estimator depends on the amount of matching discrepancy and the type of model misspecifications. 

We now compare DRE \citep{chen2020doubly}, DSM, and de-biased DSM in terms of the population mean estimates. The nonprobability sample ($S_A$) has approximately 28\% relative bias. Table \ref{tab:linearsim} shows all estimators achieve double robustness. Although there is no substantial difference between DRE and DSM estimators, DRE achieves better precision. The next section compares the performance of estimators when the implicit linearity assumption of score models is incorrect. 

\input{tables/population_estimates/linearsim.tex}

\subsection{Simulation with Nonlinear Confounders} \label{sec:nonlinear-sim}

The data generating process is unknown to a researcher in practice, and the DGPs of the propensity or prognostic scores may not be, for instance, linear in confounders available to a researcher. Hence, how the estimators perform under such misspecification is a genuine concern. The existence of any nonlinear confounders allows us to explore the case when the ``correct specification of one of the two models'' assumption \ref{a:dr} fails. This section investigates how DSM performs when such misspecification occurs. 

We keep the same data generating process in the previous section and only change the variables ``available'' to the researcher. Assume that following variables available in researcher's dataset: ${\bar{x}}_1 = x_1,\, {\bar{x}}_2=x_2 ^{2},\, {\bar{x}}_3 = x_3 ^{3},$ and ${\bar{x}}_4=x_4 ^{2}$. As the true data generating process is unknown to the researcher, she uses ${\bar{x}_1},\,{\bar{x}_2},\,{\bar{x}_3},$ and ${\bar{x}_4}$ in the estimation of the outcome and propensity scores without any further specifications/adjustments. This hypothetical situation is a better reflection of a real-world application where the analyst is ignorant of the true data generating process. 

We use the same setting as in the previous simulation and compare four scenarios. Neither of the models is correct specifications of the outcome and propensity score models. Hence, assumption \ref{a:strong-ignorability} is violated, and all estimators are misspecified. In addition, we denote a model as ``False Specification'' when ${\bar{x}_3}$ is omitted, as ``True Specification'' when all variables are included in the model. In other words, all models are misspecified because of nonlinear confounders, and some models (labelled as ``False Specfication'') are misspecified because of omitted variables.

\input{tables/population_estimates/nonlinearsim.tex}

Table \ref{tab:nonlinsim} shows that the existence of nonlinear confounders inflates the bias for all estimators as expected. However, the DRE's bias increases much faster than that of DSM. We have tested this result under various settings with nonlinear confounders and have obtained the same results. As we increase the level of nonlinearity, DRE's bias further increases (See appendix section \ref{Appendix:simulation}). The DSM's bias, on the other hand, increases at a much slower rate because the matching estimator is a semiparametric method. Our results are consistent with the literature (e.g. \citet{antonelli2018doubly, long2012doubly}). 

The DSM performs slightly better than de-biased DSM here; however, this result cannot be generalized to conclude that DSM is more robust to nonlinearity than de-biased DSM. We simulated various nonlinearity settings, and de-biased DSM performed better in some (e.g., Table \ref{tab:super_nonlinearsim}). We think one should use de-biased DSM in most cases, as it has better asymptotic properties. 

\subsection{Variance Simulations}

This section validates the double robustness of the proposed wild bootstrapping method. De-biased DSM is used to construct the confidence intervals for the sample $S_B$ mean and population mean estimates when the DGP is linear as in section \ref{sec:linearvar}. The simulations show the performance of the variance estimator for various sample sizes and the number of matchings ($M$). The performance criteria are the probability that sample and populations means are in the estimated $95\% $ confidence interval. It is computed as follows:

$$\dfrac{\sum _{s=1} ^S I\left (\mu _B ^{d,(s)} (\hat \theta) - \hat q_{0.975}<\mu _B ^{(s)}< \mu _B  ^{d,(s)}(\hat \theta) - \hat q_{0.025}\right )}{S}, $$

for $\mu _B$, and

$$\dfrac{\sum _{s=1} ^S I\left (\mu_{DSM} ^{d,(s)} (\hat \theta) - \hat q_{0.975}<\mu  ^{(s)}< \mu_{DSM} ^{d,(s)}(\hat \theta) - \hat q_{0.025}\right )}{S}, $$

for $\mu $.

\input{tables/variance/variancetable.tex}

Table \ref{tab:variancesim} shows that the wild bootstrapping performs well, and confidence intervals converge to 95\% at various sample sizes and the number of matchings. The wild bootstrapping is robust as long as one of the two models is correctly specified. Indeed our bootstrapping method ignores (1) the uncertainty of score estimates and (2) sample design (weight) for the population mean confidence intervals. Our simulations indicate that the uncertainty of the scores requires a negligible small adjustment, despite we use relatively small sample sizes and select $\rho $ small (See section \ref{sec:linearvar}). The required adjustment will be even smaller as sample sizes grow. We expect DSM mass-imputation to be used considerably larger samples (e.g., big data). 

The impact of the second problem, however, is not clear. It is known that the naive bootstrapping method could fail when samples have unequal weights \citep{barbiero2010bootstrap}. The wild bootstrapping generates valid confidence intervals in our simulations, but it may not hold in general. We leave this problem as a future research.

\section{Conclusion} \label{Conclusions}

Data integration methods are becoming increasingly popular to enhance the research prospects of new types of datasets. We propose to use DSM to do mass-imputation and consider a specific use of it: Making inference with a nonprobability sample. Our methodology is doubly robust in the sense that as long as one of the two modeling assumptions is correct, mass-imputation is asymptotically unbiased. 

Mass-imputation methods create a complete dataset, hence making any statistical analysis is straightforward with the imputed dataset. We also showed that DSM inflates the bias slower than weighting methods when models are misspecified, i.e., the data generating process possesses some degree of nonlinearity unknown to the researcher. Also, DSM is less sensitive to extreme propensity and prognostic scores because matching estimators can only interpolate the data.

While DSM has good robustness properties, it is challenging to work on its asymptotic properties. We provide consistency results and construct the confidence intervals using the residual bootstrapping approach. An essential feature of the proposed bootstrapping method is the low computational burden as it does not requires re-sampling. Our simulations show that wild bootstrapping performs well under various relative sample sizes and the number of matchings.

\newpage

\bibliographystyle{apacite}
\bibliography{ref}
\newpage 
\appendix
% Restart equation and table numbering!
\setcounter{equation}{0}
\renewcommand\theequation{A.\arabic{equation}}
\setcounter{table}{0}
\renewcommand\thetable{A.\arabic{table}}

\section{Proofs}

\subsection{Lemma 2}

\begin{proof}
    We decompose $\tilde y _i - E[y_i|\bm z_i]$ as follows:

    \begin{align}
        \tilde y _i - E[y_i|\bm z_i] &= \dfrac{1}{M} \sum _{m=1}^M y_{j_m(i;\tilde \theta)} - E[y_i|\bm z_i]\\
        &= \dfrac{1}{M} \sum _{m=1}^M \left (E[y_i|\bm z_{j_m(i;\tilde \theta)}] + \epsilon _{j_m(i;\tilde \theta)}\right ) - E[y_i|\bm z_i]\\
        &= \dfrac{1}{M} \sum _{m=1}^M \left (E[y_i|\bm z_{j_m(i;\tilde \theta)}]  - E[y_i|\bm z_i] \right ) + \dfrac{1}{M} \sum _{m=1}^M \epsilon _{j_m(i;\tilde \theta)}
    \end{align}

    $\epsilon _i$ is the error term, i.e. $y_i = E[y_i|\bm z_i] + \epsilon _i$, such that $E[\epsilon _i] = 0$. Therefore, the expected value of the second component is zero. 
    
    \begin{equation}
        E\left [ \dfrac{1}{M} \sum _{m=1}^M \epsilon _{j_m(i;\tilde \theta)} \right] = 0.
    \end{equation}
    
    The expected value of the first component is not zero for finite samples. But it converges to zero as $N_A \to \infty$. It follows from the fact that $\bm z_{j_m(i)} - \bm z_i\to 0$ for all $m=\{ 1,2,...,M\}$ because $M$ is fixed while $N_A\to \infty$. By Portmanteau lemma and assumption 4, we have $E[ y_i|\bm z_{j_m(i)}] - E[y_i|\bm z_i]\to 0$ for all $m=\{ 1,2,...,M\}$. 

    \begin{align*}
        E \left [ \tilde y _i - E[y_i|\bm z_i] \right ]&= E \left [\dfrac{1}{M} \sum _{m=1}^M \left (E[y_i|\bm z_{j_m(i;\tilde \theta)}]  - E[y_i|\bm z_i] \right ) + \dfrac{1}{M} \sum _{m=1}^M \epsilon _{j_m(i;\tilde \theta)}\right ]\\
        &= E \left [\dfrac{1}{M} \sum _{m=1}^M \left (E[y_i|\bm z_{j_m(i;\tilde \theta)}]  - E[y_i|\bm z_i] \right )\right ]\\
        &= 0
    \end{align*}
   
\end{proof}

\subsection{Lemma 3}

Before showing equation \eqref{eq:atet} is equivalent of Theorem 2 in \citet{abadie2006large} we introduce the notation for the average treatment effect for the treated estimator. Let the true ATET be $\tau $ and be ATET estimator be $\tilde \tau $:

\begin{align*}
    \tau _0&= E\big [ E[y|\bm x,W=1] - E[y|\bm x, W=0]|W=1\big],\\
    \tilde \tau_0 &= \dfrac{1}{N_B} \sum _{i\in S_B \cup S_A} W_i\{ y_i - \bar \mu _0 (\bm x_i)\} ,
\end{align*}

where $\bar \mu _0(\bm x) = E[y(0) | \bm x]$, $W$ is the treatment indicator and $y(0)$ is the potential outcome variable conditional the control group. Here, $W_i = 1$ if $i\in S_B$ and $W_i = 0$ otherwise. In other words, we consider probability sample as treatment group and nonprobability sample as control group. 

Instead of using covariate $\bm x$ we use the \textit{known} balance scores described in section 2. \citet{rosenbaum1983central} show that the covariates $\bar x$ can be substituted with estimated balance score under certain conditions. 

\begin{align}
    \tau &= E\Bigg [ E[y|\bm z, W=1] - E[y|\bm z, W=0]|W=1\Bigg ].\\
    \tilde \tau &= \dfrac{1}{N_B} \sum _{i\in S_B} \left \{ y_i - \dfrac{1}{M} \sum _{m=1} ^M y_{j_m(i;\tilde \theta)} \right \} .
\end{align}

By lemma \ref{lemma:dr} as long as one of the balance scores is a correct specification (or satisfies ignorability), then DSM must be a consistent estimator. 

Given that we rewrite the ATET in \citet{abadie2006large} with know scores and decompose $\tilde \tau - \tau$ as follows:

\begin{equation}
    \tilde \tau - \tau = (\overline{\tau(Z)} - \tau) + E_M^t + B_M^t ,\label{eq:atet_orig}
\end{equation}

where 

\begin{align}
    \overline{\tau(X)}&=\frac{1}{N_{B}} \sum_{i=1}^{N} W_{i}\left(\bar \mu\left(\bm z_{i}, 1\right)-\bar \mu_{0}\left(\bm z_{i}\right)\right), \\
    E_{M}^t&=\frac{1}{N_{B}} \sum_{i=1}^{N}\left(W_{i}-\left(1-W_{i}\right) \frac{K_{M}(i;\tilde \theta )}{M}\right) \epsilon_{i}, \\
    B_{M}^t&=\frac{1}{N_{B}} \sum_{i=1}^{N} W_{i} \frac{1}{M} \sum_{m=1}^{M}\left(\bar \mu_{0}\left(\bm z_{i}\right)-\bar \mu_{0}\left(z_{j_{m} (i;\tilde \theta)}\right)\right).
\end{align}

Here $\bar \mu (z_i,w) = E[y|\bm z_i, W_i = w]$ and $\bar \mu _{w} (z_i)=E[y_i(w)|\bm z_i]$ where $y_i (w)$ is potential outcome conditional on $W_i$. $\bar \mu (\bm z_i,w) = \bar \mu _w (\bm z_i)$ under assumption \ref{a:strong-ignorability}. Our modifications do not change $B_M^t$, but changes $(\overline{\tau(Z)} - \tau)$ and $E_M^t$. Let's decompose $\tilde \tau - \tau$:

\begin{align*}
    \tilde \tau - \tau &=  \dfrac{1}{N_B} \sum _{i\in S_B} \left \{ y_i - \dfrac{1}{M} \sum _{m=1} ^M y_{j_m(i;\tilde \theta)} \right \} - E\Big [ \bar \mu (\bm z_i, 1) - \bar \mu (\bm z_i,0)|W=1\Big ]\\
    &=- \dfrac{1}{N_B} \sum _{i\in S_B} \dfrac{1}{M} \sum _{m=1} ^M y_{j_m(i;\tilde \theta)} + \dfrac{1}{N_B} \sum _{i\in S_B}  y_i - E\Big [ \bar \mu (\bm z_i, 1) - \bar \mu (\bm z_i,0)|W=1\Big ]\\
    &=- \dfrac{1}{N_B} \sum _{i\in S_B} \dfrac{1}{M} \sum _{m=1} ^M y_{j_m(i;\tilde \theta)} + \dfrac{1}{N_B} \sum _{i\in S_B}  y_i - 0\\
    &=- \dfrac{1}{N_B} \sum _{i\in S_B} \dfrac{1}{M} \sum _{m=1} ^M y_{j_m(i;\tilde \theta)} + \dfrac{1}{N_B} \sum _{i\in S_B}  (\bar \mu _1 (z_i) + \epsilon _i)  \\
    &=- \dfrac{1}{N_B} \sum _{i\in S_B} \dfrac{1}{M} \sum _{m=1} ^M y_{j_m(i;\tilde \theta)} + \left ( \mu _B + \dfrac{1}{N_B} \sum _{i\in S_B}  \epsilon _i \right )\\
    &=- \dfrac{1}{N_B} \sum _{i\in S_B} \dfrac{1}{M} \sum _{m=1} ^M y_{j_m(i;\tilde \theta)} + \mu _B + \dfrac{1}{N_B} \sum _{i\in S_B}  \epsilon _i \\
    &= - \mu _B(\tilde \theta ) + \mu _B + \dfrac{1}{N_B} \sum _{i\in S_B}  \epsilon _i \\
    &= - \mu _B(\tilde \theta ) + \mu _B + \dfrac{1}{N_B} \sum _{i\in S_B}  \epsilon _i
\end{align*}

The decomposition exploits the fact the true ATET is known to be zero. So we expect $\tau = 0 $ because there is no treatment in the probability sample. Sample selection bias may lead to $\tau \not = 0$ but this would contradict assumption \ref{a:strong-ignorability} (ignorability) or would contradict assumption \ref{a:dr} (one of the two score models is specified correctly). In conclusion, the difference between asymptotics of ATET and DSM mass-imputation is the the noise term, $\frac{1}{N_B} \sum _{i\in S_B}  \epsilon _i$, which is taken into account in equation \eqref{eq:res}. 

So it must be:

\begin{align*}
    \mu _B (\tilde \theta) -  \mu _B &= - (\hat \tau - \tau ) + \dfrac{1}{N_B} \sum _{i\in S_B}  \epsilon _i=  \begin{cases}
        o_p(N^{-a/2})& \text{if } a\leq1,\\
        O_p(N^{-1/2})  & \text{if }  a>1.
    \end{cases}
\end{align*}

\subsection{Theorem 1}

We use the proof method in Theorem 2 of \citet{antonelli2018doubly} with some adjustments in assumptions and steps. The proof will require the following definition and results. Note that the subscript $B$ in $\mu _B$ (and in $\mu _B (\hat \theta )$) is not used in the proof for simplicity, i.e., $\mu _B = \mu $  and $\mu _B (\hat \theta ) = \mu  (\hat \theta )$.  

\textbf{Definition:} Let $U_1,\dots,U_n$ be an \textit{iid} sample with cumulative distribution function (CDF) and probability distribution function (PDF) denoted by $F$ and $f$, respectively. Then, the PDF of consecutive order statistics is given by:

\begin{equation*}
    f_{U_{(m)},U_{(m+1)}}(x,y) = b_{nm} F(x)^{m-1} (1-F(y))^{n-m-1}f(x)f(y),
\end{equation*}

with $b_{nm} = \frac{n!}{(m-1)!(n-m-1)!}$.

\textbf{Result 1:} Let $U_1,\dots,U_n$ be an \textit{iid} sample with CDF and PDF denoted by $F$ and $f$, respectively. Then, CDF of the difference of consecutive order statistics is bounded by:

\begin{equation*}
    F_{U_{(m+1)}-U_{(m)}}(x,y) \leq \int _{-\infty} ^ \infty b_{N_B M} f(x) [F(x+u)- F(x)]\partial x.
\end{equation*}

The proof of the result 1 is established in \citet{antonelli2018doubly}. 

\textbf{Result 2:} For $K>0$ and some positive random variables $\{ A_i \} _{i=1,...,N}$:

$$Pr \left (  \sum _{i=1} ^N A_i \ge K \right) \leq Pr \left ( \bigcup _{i=1} ^N \{ A_i \ge K/N \} \right) \leq \sum _{i=1} ^N Pr( A_i \ge K/N), $$

it follows from the fact that:

$$ \left \{ \sum _{i=1} ^N A_i \ge K \right \} \subset \bigcup _{i=1} ^N \{ A_i \ge K/N \} .$$

In addition to assumptions outlined in section 2:

\begin{enumerate}
    \item Matching scores are estimated on a sample that is independent of the sample used for estimation. 
    \item Necessary regularity conditions for propensity and prognostic score models are satisfied. 
    \item The distributions of the matching discrepancies for both known and estimated scores are continuous with the bounded second moments of the underlying PDFs, i.e., $\int f^2 _{D;i.\tilde{ \theta }}(x) dx < \infty $ and $\int f^2 _{D;i.\hat{ \theta }}(x) dx < \infty$.
    \item $E\left [ H_{ij} (\tilde \theta )\right]<\infty$ where: 
    
    \begin{align*}
        H_{ij} (\tilde \theta ) &= |y_j||y_i|\left | \dfrac{\partial}{\partial \tilde{ \theta}} l_{ij} (\tilde \theta)\right | \left | \dfrac{\partial}{\partial \tilde{ \theta}} l_{ji} (\tilde \theta)\right |.
    \end{align*}
\end{enumerate}

$D_{i(k)}(\theta)$ indicates the $k^{th}$ order statistics of $\{ D_{ij} (\theta ) = ||\bm z_j(\theta ) -\bm z_i(\theta )||: i\in S_B,\, j\in S_A\}$ for a given $i$, and $D_{ij} (\theta)$ has the density $f_{D;i,\theta}$. We also define $l_{ij} (\theta) = C_{iM} (\theta ) - ||\bm z_j(\theta ) -\bm z_i(\theta )||$ and $L_i (\theta )=\min _{j} \quad \{ l_{ij} (\theta ) \} = \dfrac{D_{i(M+1)}(\theta ) - D_{i(M)}(\theta )}{2}$ where:

\begin{equation*}
    C_{iM} (\theta ) = \dfrac{D_{i(M)}(\theta) + D_{i(M+1)}(\theta)}{2}.
\end{equation*}

We, later, use a smoothed version of the matching estimator:

\begin{align*}
    \mu _ \Phi (\theta; h) &=  \dfrac{1}{N_B} \sum _{i \in S_B} \dfrac{1}{M} \sum _{j \in S_A} \Phi _{h_N}\{ l_{ij}(\theta) \} y_j,
\end{align*}

where indicator function is replaced with a smoothing function, i.e., $\Phi _{h_N} (x) = (1+e^{-x/h_N})^{-1}$. $h_N$ is selected such that it converges faster than our estimator. Hence, it's convergence rate is negligible. The bandwidth is choosen as follows:

\begin{equation*}
    h_N = \dfrac{1}{N_B^3b_{N_B M}}.
\end{equation*}

\begin{proof} We decompose the error of estimate as follows:

\begin{align*}
    \mu (\hat \theta ) - \mu &= \left [ \mu (\hat \theta ) - \mu (\tilde{\theta })\right ]+ \left [\mu (\tilde \theta) - \mu \right]\\
    &= \left [ \mu (\hat \theta ) - \mu _ {\Phi } (\hat \theta; h_N ) \right ] + \left [ \mu _ {\Phi } (\hat \theta; h_N ) - \mu _ {\Phi } (\tilde \theta; h_N ) \right ] + \left [ \mu _ {\Phi } (\tilde \theta; h_N ) - \mu (\tilde \theta ) \right ] + \left [\mu (\tilde \theta) - \mu \right]\\
\end{align*}

We discuss the convergence rate of each component enumerated as follows:

\begin{enumerate}
    \item $\mu (\hat \theta ) - \mu _ {\Phi } (\hat \theta; h_N )$,
    \item $\mu _ {\Phi } (\hat \theta; h_N ) - \mu _ {\Phi } (\tilde \theta; h_N )$,
    \item $\mu _ {\Phi } (\tilde \theta; h_N ) - \mu (\tilde \theta ) $,
    \item $\mu (\tilde \theta) - \mu$.
\end{enumerate}

\textit{Proof of \# 1 and \# 2}: The difference first and second component converges to zero faster than any polynomial. Specifically, matching estimator and its smoothed version converges at rate $o_p(N_B^{-Q})$ for a given $Q\ge 0$. We will show the convergence of the first one which can be directly applied to the second. Rewriting the term:

\begin{align*}
    \mu (\theta ) &=  \dfrac{1}{N_B} \sum _{i \in S_B} \dfrac{1}{M} \sum _{j \in S_A} I \{ l_{ij}(\theta)>0 \} y_j\   
\end{align*}

Then, 

\begin{align}
    |\mu _ \Phi (\theta; h_N) - \mu (\theta )| &= \left|  \dfrac{1}{N_B M} \sum _{i \in S_B} \sum _{j \in S_A} (\Phi _{h_N} (l_{ij} (\theta )) - I \{ l_{ij}(\theta)>0 \} )  \right|\\
    &\leq   \dfrac{1}{ M} \sqrt{\dfrac{1}{N_B^2}\sum _{i \in S_B} \sum _{j \in S_A} y_j ^2} \sqrt{\sum _{i \in S_B} \sum _{j \in S_A} (\Phi _{h_N} (l_{ij} (\theta )) - I \{ l_{ij}(\theta)>0 \} )^2}  \\
    &= \dfrac{1}{ M} \sqrt{\dfrac{1}{N_B} \sum _{j \in S_A} y_j ^2} \sqrt{\sum _{i \in S_B} \sum _{j \in S_A} (\Phi _{h_N} (l_{ij} (\theta )) - I \{ l_{ij}(\theta)>0 \} )^2}  \\
    &= O_p(1) \sqrt{\sum _{i \in S_B} \sum _{j \in S_A} (\Phi _{h_N} (l_{ij} (\theta )) - I \{ l_{ij}(\theta)>0 \} )^2}  
\end{align}

We use Cauchy-Schwarz inequality in the first step and use the assumption that $a=1$ assumption in the last step, i.e., $\sqrt{\dfrac{1}{N_B} \sum _{j \in S_A} y_j ^2}=O_p(1)$. Note that we sum over sample $A$ and divide with sample size $N_B$. Our reasoning is justified if $a=1$, in other words, two samples converge to infinity with the same rate. 

We, now, focus on $\Phi _{h_N} (l_{ij} (\theta )) - I \{ l_{ij}(\theta)>0 \}$ and rewrite it as follows:

\begin{equation}
    \Phi _{h_N} (l_{ij} (\theta )) - I \{ l_{ij}(\theta)>0 \} = \dfrac{\text{sign} \{ - l_{ij} (\theta ) \} }{e^{|l_{ij}(\theta )|/h_N } + 1}
\end{equation}

Having $\theta = \hat \theta $:

\begin{align}
    |\mu _ \Phi (\hat \theta; h_N) - \mu (\hat \theta )| &\leq  O_p(1) \sqrt{\sum _{i \in S_B} \sum _{j \in S_A} \left ( \dfrac{\text{sign} \{ - l_{ij} (\hat \theta ) \} }{e^{|l_{ij}(\hat \theta )|/h_N } + 1} \right )^2}  \\
    &=  O_p(1) \sqrt{\sum _{i \in S_B} \sum _{j \in S_A}  \dfrac{1}{\left (e^{|l_{ij}(\hat \theta )|/h_N } + 1\right )^2} } \\
    &\leq  O_p(1) \sqrt{\sum _{i \in S_B}  \dfrac{N_A}{\left (e^{|L_{i}(\hat \theta )|/h_N } + 1\right )^2} } \\
    &\leq  O_p(1) \sqrt{\sum _{i \in S_B}  \dfrac{N_A}{ e^{L_{i}(\hat \theta )/h_N }} } 
\end{align}

Following from result 2, for any $0<K\leq 1 $ and $Q>0$:

\begin{align}
    \lim _{N_B \to \infty} & Pr \left \{  \sum _{i\in S_B} \dfrac{N_B ^{Q} N_A}{e^{L_{i}(\hat \theta )/h_N}} > K \right \} \leq \lim _{N_B \to \infty}  \sum _{i\in S_B} Pr \left \{  \dfrac{N_B ^{Q} N_A}{e^{L_{i}(\hat \theta )/h_N}} > \dfrac{K}{N_B} \right \} \\
    &\leq \lim _{N_B \to \infty}  \sum _{i\in S_B} Pr \left \{ L_{i}(\hat \theta )/h_N < -\log K + \log N_A + (Q+1)\log N_B \right \} \\
    &\leq \lim _{N_B \to \infty}  \sum _{i\in S_B} Pr \left \{ L_{i}(\hat \theta ) < \dfrac{-\log K + \log N_A + (Q+1)\log N_B}{N_B ^3 b_{N_B M}} \right \} \\
    &= \lim _{N_B \to \infty}  \sum _{i\in S_B} Pr \left \{ \dfrac{D_{i(M+1)}(\theta ) - D_{i(M)}(\theta )}{2} < \dfrac{-\log K + \log N_A + (Q+1)\log N_B}{N_B ^3 b_{N_B M}} \right \} 
\end{align}

We take log of the inequality and do some algebra. The next steps uses the definitions of $h_N$ and $L_i (\hat \theta)$, respectively. The last equation and the result 1 imply that:

\begin{align}
    \lim _{N_B \to \infty} & Pr \left \{  \sum _{i\in S_B} \dfrac{N_B ^{Q} N_A}{e^{L_{i}(\hat \theta )/h_N}} > K \right \} \\
    &\leq  \lim _{N_B \to \infty}  \sum _{i\in S_B} \int _{-\infty } ^ \infty b_{N_B M} f_i (x) \left [ F_i \left ( x +  \dfrac{-2\log K + 2\log N_A + 2(Q+1)\log N_B}{N_B ^3 b_{N_B M}} \right ) - F_i (x)\right ] \partial x 
\end{align}

$F_i$ and $f_i$ are CDF and PDF, respectively, of $D_{ij} (\hat \theta )$ for given $i$. The use of result 1 is justified by the assumption that $\hat \theta $ is estimated on a sample that is independent from the estimation sample, and hence $D_{ij} (\hat \theta ) $ are independent for given $i$. 

We use mean value theorem and expand $F_i (x+u)$ as follows: 

$$ 
F_i (x+u) = F_i (x) + u.f_i(x^*) \qquad \text{where} \qquad x^*\in [x,x+u]
$$

Hence,

\begin{align}
    \lim _{N_B \to \infty} & Pr \left \{  \sum _{i\in S_B} \dfrac{N_B ^{Q} N_A}{e^{L_{i}(\hat \theta )/h_N}} > K \right \} \\
    &\leq  \lim _{N_B \to \infty}  \sum _{i\in S_B} \int _{-\infty } ^ \infty b_{N_B M} f_i (x) \left [  \left ( \dfrac{-2\log K + 2\log N_A + 2(Q+1)\log N_B}{N_B ^3 b_{N_B M}} \right )f_i (x^*)\right ] \partial x \\
    &\leq  \lim _{N_B \to \infty}  \sum _{i\in S_B}   \dfrac{-2\log K + 2\log N_A + 2(Q+1)\log N_B}{N_B ^3 }  \int _{-\infty } ^ \infty f_i ^2 (x) \partial x \\
    &\leq   \lim _{N_B \to \infty} N_B \dfrac{-2\log K + 2\log N_A + 2(Q+1)\log N_B}{N_B ^3 } C \\
    &= 0
\end{align}

because $\int _{-\infty } ^ \infty f_i ^2 (x) \partial x$ is bounded and $N_A/N_B \to A,\, A\in (0,\infty )$ by assumption \ref{a:relative-sample-size}. Consequently, 

\begin{align*}
    |\mu _ \Phi (\hat \theta; h_N) - \mu (\hat \theta )| &\leq O_p(1)o_p(N^{-Q})\\
    &=o_p(N^{-Q})
\end{align*}

for any $Q\ge 0$. The same proof method applies for $|\mu _ \Phi (\tilde \theta; h_N) - \mu ( \tilde \theta )|$. 

\textit{Proof of \# 3:} The Taylor expansion of $\mu _ {\Phi } (\hat \theta; h_N ) - \mu _ {\Phi } (\tilde \theta; h_N )$ yields:

$$
\mu _ {\Phi } (\hat \theta; h_N ) - \mu _ {\Phi } (\tilde \theta; h_N ) =  \dfrac{\partial}{\partial \tilde \theta } \mu _ {\Phi } (\tilde \theta; h_N )'(\hat \theta - \tilde \theta ) + O_p \left (||\hat \theta - \tilde \theta|| ^2\right )
$$

We further decompose $\dfrac{\partial}{\partial \tilde \theta } \mu _ {\Phi } (\tilde \theta; h_N )$ as follows:

\begin{align}
    \left | \dfrac{\partial}{\partial \tilde \theta } \mu _ {\Phi } (\tilde \theta; h_N ) \right | &= \left | \dfrac{\partial}{\partial \tilde \theta } \dfrac{1}{N_B} \sum _{i\in S_B} \dfrac{1}{M} \sum _{j \in S_A} \Phi _{h_N} (l_{ij} (\tilde \theta))y_j \right |\\
    &= \left |  \dfrac{1}{M N_B} \sum _{i\in S_B}  \sum _{j \in S_A} \phi _{h_N} (l_{ij} (\tilde \theta))y_j \dfrac{\partial}{\partial \tilde \theta } l_{ij} (\tilde \theta)  \right |\\
    &\leq \left |  \dfrac{1}{M N_B} \sum _{i\in S_B} \phi _{h_N} (L_{i} (\tilde \theta))  \sum _{j \in S_A} y_j \dfrac{\partial}{\partial \tilde \theta } l_{ij} (\tilde \theta)  \right |\\
    &= \dfrac{1}{M N_B} \sum _{i\in S_B} \phi _{h_N} (L_{i} (\tilde \theta))  \sum _{j \in S_A} \left | y_j \dfrac{\partial}{\partial \tilde \theta } l_{ij} (\tilde \theta)  \right |\\
    &\leq \dfrac{1}{M N_B} \sqrt{\sum _{i\in S_B} \phi _{h_N} ^2 (L_{i} (\tilde \theta)) }  \sqrt{\sum _{i\in S_B} \left ( \sum _{j \in S_A} \left | y_j \dfrac{\partial}{\partial \tilde \theta } l_{ij} (\tilde \theta)  \right | \right) ^2}\\
    &= \dfrac{1}{M } \sqrt{N_B \sum _{i\in S_B} \phi _{h_N} ^2 (L_{i} (\tilde \theta)) }  \sqrt{N_B ^{-3}\sum _{i\in S_B} \left ( \sum _{j \in S_A} \left | y_j \dfrac{\partial}{\partial \tilde \theta } l_{ij} (\tilde \theta)  \right | \right) ^2}\\
    &= \dfrac{1}{M } \sqrt{N_B \sum _{i\in S_B} \phi _{h_N} ^2 (L_{i} (\tilde \theta)) }  \sqrt{N_B ^{-3}\sum _{i\in S_B}  \sum _{j \in S_A} \sum _{k \in S_A}  |y_j||y_k| \left | \dfrac{\partial}{\partial \tilde \theta } l_{ij} (\tilde \theta) \right | \left | \dfrac{\partial}{\partial \tilde \theta } l_{ik} (\tilde \theta)  \right |}\\
    &\leq \dfrac{1}{M } \sqrt{N_B \sum _{i\in S_B} \phi _{h_N} ^2 (L_{i} (\tilde \theta)) }  \sqrt{N_B ^{-3}\sum _{i\in S_B}  \sum _{j \in S_A} \sum _{k \in S_A}  H}\\
    &= \dfrac{1}{M } \sqrt{N_B \sum _{i\in S_B} \phi _{h_N} ^2 (L_{i} (\tilde \theta)) }  \sqrt{N_B ^{-2} N_A ^2  H}\\
    &\leq \dfrac{1}{M } \sqrt{N_B \sum _{i\in S_B} \phi _{h_N} ^2 (L_{i} (\tilde \theta)) } \,\, O_p(1)
\end{align}

where $\phi _{h_N} (x) = \partial \Phi _{h_N} (x)/ \partial x = \frac{e^{x/h_N}}{h_N (e^{x/h_N} + 1)^2}$ and,

$$ 
H = \max _{i\in S_B,\, j\in S_A,\, k\in S_A} \quad |y_j||y_k| \left | \dfrac{\partial}{\partial \tilde \theta } l_{ij} (\tilde \theta) \right | \left | \dfrac{\partial}{\partial \tilde \theta } l_{ik} (\tilde \theta)  \right | \qquad \text{where} \qquad 0<H<\infty.
$$

by assumption of boundedness. 

For any $0<K\leq 1$, 

$$
\lim _{N_B \to \infty } Pr \left \{ N_B \sum _{i\in S_B } \phi ^2 _{h_N} (L_i (\tilde \theta )) > K \right \} = 0 
$$

with the similar arguments given in proof of \# 1 and \# 2. 

\begin{align}
    \lim _{N_B \to \infty } Pr &\left \{ N_B \sum _{i\in S_B } \phi ^2 _{h_N} (L_i (\tilde \theta )) > K \right \} = \lim _{N_B \to \infty } Pr \left \{ N_B \sum _{i\in S_B } \left ( \dfrac{e ^{L_i (\tilde \theta )/h_N}}{h_N(1 + e^{L_i (\tilde \theta )/h_N })^2}\right )^2 > K \right \} \\
    &\leq  \lim _{N_B \to \infty } \sum _{i\in S_B } Pr \left \{ N_B  \left ( \dfrac{e ^{L_i (\tilde \theta )/h_N}}{h_N(1 + e^{L_i (\tilde \theta )/h_N })^2}\right )^2 > K \right \} \\
    &\leq  \lim _{N_B \to \infty } \sum _{i\in S_B } Pr \left \{ N_B  \left ( \dfrac{1+e ^{L_i (\tilde \theta )/h_N}}{h_N(1 + e^{L_i (\tilde \theta )/h_N })^2}\right )^2 > K \right \} \\
    &=  \lim _{N_B \to \infty } \sum _{i\in S_B } Pr \left \{ N_B  \left ( \dfrac{1}{h_N(1 + e^{L_i (\tilde \theta )/h_N })}\right )^2 > K \right \} \\
    &\leq  \lim _{N_B \to \infty } \sum _{i\in S_B } Pr \left \{ N_B  \left ( \dfrac{1}{h_N( e^{L_i (\tilde \theta )/h_N })}\right )^2 > K \right \} \\
    &=  \lim _{N_B \to \infty } \sum _{i\in S_B } Pr \left \{\log N_B -  2 \left  ( \log h_N + L_i (\tilde \theta )/h_N )\right ) > \log K \right \} \\
    &=  \lim _{N_B \to \infty } \sum _{i\in S_B } Pr \left \{L_i (\tilde \theta ) < \dfrac{\log N_B - \log K + 6\log N_B + \log b_{N_B M}}{2N_B ^3 b_{N_B M}} \right \} \\
    &=  \lim _{N_B \to \infty } \sum _{i\in S_B } Pr \left \{D_{i(M+1)}(\theta ) - D_{i(M)}(\theta ) < \dfrac{- \log K + 7\log N_B + \log b_{N_B M}}{N_B ^3 b_{N_B M}} \right \} \\
    &\leq \lim _{N_B \to \infty } \sum _{i\in S_B } \int _{-\infty } ^ \infty b_{N_B M} f_i (x) \left [ F_i \left ( x +  \dfrac{- \log K + 7\log N_B + \log b_{N_B M}}{N_B ^3 b_{N_B M}} \right ) - F_i (x)\right ] \partial x 
\end{align}

Using the Results 1 and 2 along with the mean value theorem as previously, $F_i (x+u) = F_i (x) + uf_i(x^*)$ where $x^*\in [x,x+u]$, we obtain the following:

\begin{align}
    \lim _{N_B \to \infty } Pr &\left \{ N_B \sum _{i\in S_B } \phi ^2 _{h_N} (L_i (\tilde \theta )) > K \right \} \\
    &\leq \lim _{N_B \to \infty}  \sum _{i\in S_B} \int _{-\infty } ^ \infty b_{N_B M} f_i (x) \left [  \left ( \dfrac{- \log K +7 \log N_B + \log b_{N_B M}}{N_B ^3 b_{N_B M}} \right )f_i (x^*)\right ] \partial x \\
    &\leq \lim _{N_B \to \infty}  \sum _{i\in S_B}   \dfrac{- \log K +7 \log N_B + \log b_{N_B M}}{N_B ^3} \int _{-\infty } ^ \infty f_i^2 (x) \partial x \\
    &\leq \lim _{N_B \to \infty}  \sum _{i\in S_B}   \dfrac{- \log K +7 \log N_B + \log b_{N_B M}}{N_B ^3} C\\
    &= \lim _{N_B \to \infty}  \sum _{i\in S_B}   \dfrac{- \log K + 5\log N_B + \log \left ( \dfrac{N_B!}{(M-1)!(N_B - M - 1)!}\right )}{N_B ^3} C\\
    & \leq \lim _{N_B \to \infty}  \sum _{i\in S_B}   \dfrac{- \log K + 7 \log N_B + \sum _{N=N_B - M} ^{N_B} \log N }{N_B ^3} C\\
    & \leq \lim _{N_B \to \infty}  \sum _{i\in S_B}   \dfrac{- \log K + 7\log N_B + N_B \log N_B }{N_B ^3} C\\
    & \leq \lim _{N_B \to \infty}  \dfrac{- N_B\log K - N_B5\log N_B - N_B^2 \log N_B }{N_B ^3} C\\
    &\leq 0
\end{align}

Thus,

$$
\left | \dfrac{\partial}{\partial \tilde \theta } \mu _{\Phi } (\tilde \theta ; h_N) \right | = o_p (1).
$$

Assuming that convergence rate of $\hat \theta $ to $\tilde \theta $ is $\sqrt{\dfrac{1}{N}}$, then it must be:

$$
\mu _ {\Phi } (\hat \theta; h_N ) - \mu _ {\Phi } (\tilde \theta; h_N ) = O_p (N_B ^{-1/2})
$$

\textit{Proof of \# 4}: The last step directly follows from lemma \ref{lemma:convergenceinprob}:

$$ 
\mu _B (\tilde \theta ) - \mu _B = O_p(N_B ^{-1/2})
$$.

All steps combined implies that

$$ 
\mu (\hat \theta ) - \mu = O_p(N_B^{-1/2})
$$

which completes the proof. 

\end{proof}

\subsection{Theorem 2}

\begin{proof}
    Let's decompose $\mu _{DSM} (\hat \theta) - \mu$ as follows:

    $$
    (\mu _{DSM} (\hat \theta ) - \mu _{DSM} (\tilde \theta )) + (\mu _{DSM} (\tilde \theta ) - \mu _{\Psi }) + (\mu _{\Psi } - \mu)
    $$

    The first component consists of the bias terms and error terms associated with matching outcomes. 

    \begin{align}
        \mu _{DSM} (\hat \theta ) &- \mu _{DSM} (\tilde \theta ) = \notag \\
        &\dfrac{1}{M \sum _{i\in S_B} d_i} \sum _{i\in S_B} d_i \sum _{j 
        \in J_M(i;\hat \theta ) }y_j - \dfrac{1}{M \sum _{i\in S_B} d_i} \sum _{i\in S_B} d_i \sum _{j 
        \in J_M(i;\tilde \theta ) }y_j\\
        &= \dfrac{1}{M \hat N} \sum _{i\in S_B} d_i \sum _{j 
        \in J_M(i;\hat \theta ) }y_j - \dfrac{1}{M \hat N} \sum _{i\in S_B} d_i \sum _{j 
        \in J_M(i;\tilde \theta ) }y_j\\
        &= \dfrac{1}{M \hat N} \sum _{i\in S_B} d_i \sum _{m=1} ^ M y_{j_m(i;\hat \theta )} - \dfrac{1}{M \hat N} \sum _{i\in S_B} d_i \sum _{m=1} ^M y_{j_m(i;\tilde \theta )}\\
        &= \dfrac{1}{M \hat N} \left [\sum _{i\in S_B} d_i \sum _{m 
        =1 } ^M (E[y_{j_m(i;\hat \theta )}|{ \bm{z}} _{j_m(i;\hat \theta )}] - E[y_{j_m(i;\hat \theta )}|{ \bm{z}} _{j_m(i;\tilde \theta )}]   + \epsilon _{j_m(i;\hat \theta )} - \epsilon _{j_m(i;\tilde \theta )} )\right ]
    \end{align}

    As the error terms are assumed to be uncorrelated with the sample weights, they will converge at rate $O_p(N_B ^{-1/2}) $:

    \begin{align}
        \mu _{DSM} (\hat \theta ) &- \mu _{DSM} (\tilde \theta ) = \\
        & \dfrac{1}{M \hat N} \left [\sum _{i\in S_B} d_i \sum _{m 
        =1 } ^M (E[y_{j_m(i;\hat \theta )}|{ \bm{z}} _{j_m(i;\hat \theta )}] - E[y_{j_m(i;\hat \theta )}|{ \bm{z}} _{j_m(i;\tilde \theta )}])\right ] + O_p(N_B ^{-1/2}).
    \end{align}

    The remaining one is the bias term that is known converging at rate $O_p(N_B^{-1/2})$. Hence,

    \begin{equation}
        \mu _{DSM} (\hat \theta ) - \mu _{DSM} (\tilde \theta ) = O_p(N_B ^{-1/2})
    \end{equation}

    The second component consists of the difference between imputed $y_i$ estimated with $\tilde \theta $ and known $y_i$. The difference between two terms are equal to error terms because their conditional means are equal. It follows from that fact that $\bm z_{j_m(i;\tilde \theta )} \to \bm  z _i$, and $E[y_i| \bm z_{j_m(i;\tilde \theta )}] = E[y_i|\bm z_i]$ as  $N_B\to \infty$  (and as $N_A \to \infty)$. 

    \begin{align}
        \mu _{DSM} (\tilde \theta ) - \mu _{\Psi } &= \dfrac{1}{M \hat N} \sum _{i\in S_B} d_i \sum _{j 
        \in J_M(i;\tilde \theta ) }y_j - \dfrac{1}{M \hat N} \sum _{i\in S_B} d_iy_i \\
        &= \dfrac{1}{M \hat N} \sum _{i\in S_B} d_i \sum _{j 
        \in J_M(i;\tilde \theta ) }(y_j - y_i) \\
        &= \dfrac{1}{M \hat N} \sum _{i\in S_B} d_i \sum _{j 
        \in J_M(i;\tilde \theta ) }(\epsilon _j - \epsilon _i) \\
        &= O_p(N_B ^{-1/2})
    \end{align}
    The last component $(\mu _{\Psi } - \mu)$ is the well-known Hajek Estimator \citep{hajek1964asymptotic}. Consequently, 

    \begin{equation}
        \mu _{DSM} (\hat \theta ) -\mu = O_p(N_B ^{-1/2}).
    \end{equation}

\end{proof}

\newpage 

\section{Simulations}\label{Appendix:simulation}

This simulation replicates the linear simulation except that it assumes researchers observes the following variables ${\bar{x}}_1 = x_1,\, {\bar{x}}_2=x_2 ^{1.15},\, {\bar{x}}_3 = x_3 ^{-0.85},$, and ${\bar{x}}_4=x_4 ^{-1.15}$. 

We observe that under such extreme nonlinearization settings, DSM still performs well - that it is substantially less sensitive to functional form misspecifications. 

\input{tables/population_estimates/extremenonlinear.tex}

\end{document}

%% file: tables/sample S_B table/sampleb.tex
% latex table generated in R 3.6.2 by xtable 1.8-4 package
% Sat Aug 07 23:12:13 2021
\begin{table}[!ht]
\centering
\caption{Simulation Results of Probability Sample Mean Estimates } 
\label{tab:sampleB}
\vspace{5pt}
\begin{tabular}{l|rrr}
  \hline\hline 
 & \textbf{Mean} & \textbf{RB} & \textbf{MSE} \\ 
  \hline
Sample $S_B$ & 10.120 & 0.000 & 0.000 \\ 
  DSM (TT) & 10.080 & -0.390 & 0.392 \\ 
  DSM (FT) & 10.160 & 0.395 & 0.406 \\ 
  DSM (TF) & 10.136 & 0.156 & 0.390 \\ 
  DSM (FF) & 12.358 & 22.126 & 5.406 \\ 
  De-Biased DSM (TT) & 10.100 & -0.193 & 0.399 \\ 
  De-Biased DSM (FT) & 10.189 & 0.688 & 0.414 \\ 
  De-Biased DSM (TF) & 10.111 & -0.086 & 0.393 \\ 
  De-Biased DSM (FF) & 12.386 & 22.403 & 5.542 \\ 
   \hline
\end{tabular}
\end{table}

%% file: tables/population_estimates/linearsim.tex
% latex table generated in R 3.6.2 by xtable 1.8-4 package
% Sat Aug 07 23:12:25 2021
\begin{table}[!hb]
\centering
\caption{Simulation Results with Linear Confounders} 
\label{tab:linearsim}
\vspace{5pt}
\begin{tabular}{l|rrr}
  \hline\hline 
 & \textbf{Mean} & \textbf{RB} & \textbf{MSE} \\ 
  \hline
Population Mean & 9.278 & 0.000 & 0.000 \\ 
  Sample $A$ Mean & 11.906 & 28.331 & 6.949 \\ 
  DRE (TT) & 9.271 & -0.076 & 0.315 \\ 
  DRE (FT) & 9.254 & -0.262 & 0.347 \\ 
  DRE (TF) & 9.270 & -0.087 & 0.290 \\ 
  DRE (FF) & 11.577 & 24.777 & 5.550 \\ 
  DSM (TT) & 9.268 & -0.107 & 0.406 \\ 
  DSM (FT) & 9.326 & 0.518 & 0.396 \\ 
  DSM (TF) & 9.310 & 0.345 & 0.380 \\ 
  DSM (FF) & 11.569 & 24.693 & 5.607 \\ 
  De-biased DSM (TT) & 9.268 & -0.105 & 0.407 \\ 
  De-biased DSM (FT) & 9.330 & 0.564 & 0.397 \\ 
  De-biased DSM (TF) & 9.282 & 0.044 & 0.380 \\ 
  De-biased DSM (FF) & 11.577 & 24.777 & 5.642 \\ 
   \hline
\end{tabular}
\end{table}

%% file: tables/population_estimates/nonlinearsim.tex
% latex table generated in R 3.6.2 by xtable 1.8-4 package
% Sat Aug 07 23:12:25 2021
\begin{table}[!hb]
\centering
\caption{Simulation Results with Nonlinear Confounders} 
\label{tab:nonlinsim}
\vspace{5pt}
\begin{tabular}{l|rrr}
  \hline\hline 
 & \textbf{Mean} & \textbf{RB} & \textbf{MSE} \\ 
  \hline
Population Mean & 9.278 & 0.000 & 0.000 \\ 
  Sample $A$ Mean & 11.921 & 28.484 & 7.026 \\ 
  DRE (TT) & 9.600 & 3.466 & 0.395 \\ 
  DRE (FT) & 9.819 & 5.827 & 0.621 \\ 
  DRE (TF) & 9.926 & 6.983 & 0.695 \\ 
  DRE (FF) & 11.665 & 25.724 & 5.955 \\ 
  DSM (TT) & 9.382 & 1.117 & 0.407 \\ 
  DSM (FT) & 9.386 & 1.166 & 0.412 \\ 
  DSM (TF) & 9.396 & 1.271 & 0.414 \\ 
  DSM (FF) & 11.538 & 24.359 & 5.473 \\ 
  De-biased DSM (TT) & 9.392 & 1.229 & 0.409 \\ 
  De-biased DSM (FT) & 9.400 & 1.315 & 0.416 \\ 
  De-biased DSM (TF) & 9.401 & 1.321 & 0.415 \\ 
  De-biased DSM (FF) & 11.559 & 24.586 & 5.567 \\ 
   \hline
\end{tabular}
\end{table}

%% file: tables/variance/variancetable.tex
% latex table generated in R 3.6.2 by xtable 1.8-4 package
% Sat Aug 07 23:11:54 2021
\begin{table}[ht]
\centering
\caption{Percentage of simulation estimations that are within the 95\% Confidence Interval} 
\label{tab:variancesim}
\begin{tabular}{lll|llll|llll}
 \\ \hline\hline 
 \multicolumn{1}{l}{} & \multicolumn{2}{c}{\textbf{Sample Sizes}} & \multicolumn{4}{c}{\textbf{Sample $S_B$}} & \multicolumn{4}{c}{\textbf{Population}}\\ \hline
$M$ & $N_A$ & $N_B$ & \textbf{TT} & \textbf{FT} & \textbf{TF} & \textbf{FF} & \textbf{TT} & \textbf{FT} & \textbf{TF} & \textbf{FF} \\ 
  \hline
3 & 500 & 1000 & 0.930 & 0.948 & 0.934 & 0.042 & 0.944 & 0.950 & 0.940 & 0.032 \\ 
  3 & 1000 & 500 & 0.940 & 0.951 & 0.948 & 0.023 & 0.948 & 0.952 & 0.952 & 0.017 \\ 
  5 & 1000 & 500 & 0.949 & 0.948 & 0.942 & 0.007 & 0.940 & 0.950 & 0.944 & 0.010 \\ 
  5 & 1000 & 1000 & 0.945 & 0.954 & 0.942 & 0.003 & 0.949 & 0.955 & 0.951 & 0.004 \\ 
  6 & 1000 & 2000 & 0.933 & 0.954 & 0.930 & 0.001 & 0.954 & 0.963 & 0.956 & 0.000 \\ 
  8 & 1500 & 1000 & 0.936 & 0.946 & 0.931 & 0.000 & 0.940 & 0.941 & 0.944 & 0.000 \\ 
  8 & 1500 & 1500 & 0.944 & 0.954 & 0.940 & 0.000 & 0.942 & 0.954 & 0.948 & 0.000 \\ 
  10 & 2000 & 2000 & 0.939 & 0.953 & 0.940 & 0.000 & 0.950 & 0.953 & 0.942 & 0.000 \\ 
  10 & 2500 & 2500 & 0.941 & 0.954 & 0.942 & 0.000 & 0.942 & 0.954 & 0.946 & 0.000 \\ 
  15 & 3000 & 1500 & 0.942 & 0.946 & 0.941 & 0.000 & 0.950 & 0.944 & 0.934 & 0.000 \\ 
   \hline
\end{tabular}
\end{table}

%% file: tables/population_estimates/extremenonlinear.tex
% latex table generated in R 3.6.2 by xtable 1.8-4 package
% Sat Aug 07 23:12:25 2021
\begin{table}[!hb]
\centering
\caption{Simulation Results with Nonlinear Confounders} 
\label{tab:super_nonlinearsim}
\vspace{5pt}
\begin{tabular}{l|rrr}
  \hline\hline 
 & \textbf{Mean} & \textbf{RB} & \textbf{MSE} \\ 
  \hline
Population Mean & 9.278 & 0.000 & 0.000 \\ 
  Sample $A$ Mean & 11.914 & 28.405 & 6.983 \\ 
  DRE (TT) & 19.451 & 109.640 & 177.945 \\ 
  DRE (FT) & 7.586 & -18.242 & 31.011 \\ 
  DRE (TF) & 10.426 & 12.368 & 1.817 \\ 
  DRE (FF) & 11.697 & 26.065 & 6.141 \\ 
  DSM (TT) & 9.631 & 3.803 & 0.543 \\ 
  DSM (FT) & 9.684 & 4.369 & 0.588 \\ 
  DSM (TF) & 9.832 & 5.966 & 0.753 \\ 
  DSM (FF) & 11.572 & 24.723 & 5.614 \\ 
  De-biased DSM (TT) & 9.492 & 2.303 & 0.497 \\ 
  De-biased DSM (FT) & 9.653 & 4.036 & 0.571 \\ 
  De-biased DSM (TF) & 9.598 & 3.449 & 0.614 \\ 
  De-biased DSM (FF) & 11.543 & 24.407 & 5.487 \\ 
   \hline
\end{tabular}
\end{table}